\documentclass[aps,pra,twocolumn,superscriptaddress,floatfix]{revtex4-2}

\usepackage{amsmath,amssymb}
\usepackage{graphicx}
\usepackage{booktabs}

\newcommand{\rhob}{r_{\mathrm{b}}}
\newcommand{\Nstar}{N^{*}}
\newcommand{\zenodo}{10.5281/zenodo.23993247}
\newcommand{\zenodocode}{10.5281/zenodo.24052266}
\newcommand{\zenodonote}{10.5281/zenodo.24128931}

\begin{document}

\title{Model-family selection and recovery-gate design\\
in generator-based noise tomography}

\author{Long Huang}
\email{huangl@sicnu.edu.cn}
\affiliation{College of Physics and Electronic Engineering, Sichuan Normal University,
Chengdu 610101, China}

\date{\today}

\begin{abstract}
Family selection in generator-model noise tomography is a hypothesis test whose
shot floor has closed form under a purely coherent noise model,
$\Nstar=\lambda_{\mathrm{req}}/(\rhob S)^{2}$: set by the absolute bias
$\rhob S$, not the relative residual, and computable from a calibration alone.
The same computation prices the recovery gates, fingerprint and Fisher
information coming off the same map, so there is no design-independent verdict
and no family-independent design optimum. Whether a design can move that price
is structural: a bias confined to the block every design measures identically
is design-blind, one that reaches the code blocks is repriced by a factor of
sixty-eight---from a dominated design to the cheapest, not between two
optima---and the two platforms studied here are its extremes. At the
reference point the floor is $1.1\times10^{7}$ and $6.7\times10^{6}$ shots per
component, the easy corner of the sampled working points, whose median floor is
eighty-three times dearer in one direction and less than twice in the other; at
the $10^{5}$ shots a run spends, the power equals
the level and the residual reported is shot noise. The design problem is
conditioning, not rank: across the sixty-four placements the rms varies by
factors of $2.7$ and $3.9$, while the optimal gate count is flat at two to
four. At five qubits the whole sample moves to $10^{9}$--$2\times10^{10}$ shots, a
design spread of twenty-one against nine at three qubits; on a measured
three-qubit register the floor is a requirement
on systematic control before it is a shot count, and where that control is
missing no budget supplies it.
\end{abstract}

\maketitle

\section{Introduction}
\label{sec:intro}

In the near term a quantum device is characterised not by its full process matrix
but by a small coherent generator: a handful of rotation angles on each qubit and
a handful of two-qubit coupling angles, fitted to a measured fingerprint. The
programme has a long history under several names---self-consistent process
tomography \cite{merkel}, gate set tomography \cite{gst,greenbaum}, and the
linear-inversion, direct-estimation and compressed-sensing variants that make it
affordable \cite{dasilva,flammia,harper}---and its common structure is that the
estimator is a least-squares fit of a parametric model to a fixed set of
measurable probabilities.
Reviews of the field \cite{eisert} classify these protocols by what they
promise and at what sample cost; both decisions below sit inside that cost,
and neither is usually counted in it.

Two decisions precede that fit, and both are usually made once and quoted as
facts about the apparatus. The first is the \emph{family}: which generators the
model is allowed to contain.
In the taxonomy of small Markovian errors these are reduced models, obtained
by keeping a sector of the error-generator basis and discarding the rest
\cite{taxonomy}, so the choice is a modelling decision with a statistical
price and not a label attached to a device. A fixed-frequency transmon pair carries an always-on
dispersive coupling and needs a family of the form
$\mathfrak{su}(2)^{\otimes3}\oplus\mathrm{span}\{Z_jZ_{j+1}\}$, eleven parameters
on three qubits. A tunable coupler carries the exchange
$J(X_jX_{j+1}+Y_jY_{j+1})/2$ and needs
$\mathfrak{su}(2)^{\otimes3}\oplus\mathrm{span}\{X_jX_{j+1},Y_jY_{j+1}\}$,
thirteen. The second is the \emph{recovery design}: which sites of the code get a
correcting Pauli before the next parity check, a choice that changes which
generators the fingerprint is sensitive to and therefore how well each parameter
is determined.

The two decisions are treated as independent, and each is settled by a single
number. Families are chosen by the residual of the fit, on the argument that the
correct family can represent the device exactly and the wrong one cannot. This is
a selection criterion of the usual kind \cite{burnham}, and like every such
criterion it has a sampling distribution that the bare comparison of $0$ against
$0.4\%$ hides. Designs are chosen by the gate count that minimises the
Cram\'er--Rao bound, on the argument that gates are the resource being spent.
Neither argument is wrong. Both are incomplete in the same way: each compares two
numbers that were computed at a shot budget and at a working point which the
comparison silently assumes.

This paper supplies what they leave out. The results are five.

\emph{(i) Family selection is a detection problem with a computable price.} The
residual that separates two families is a bias measured through noise, so choosing
between them is a hypothesis test with a shot floor. Because the statistic is
non-central $\chi^{2}$ with $\lambda=\rhob^{2}S^{2}N$, the floor is closed,
$\Nstar=\lambda_{\mathrm{req}}/(\rhob S)^{2}$, and $\lambda_{\mathrm{req}}$
depends on the degrees of freedom only as $\sqrt{\nu}$. What governs
detectability is the absolute bias $\rhob S$, not the fractional residual $\rhob$
usually reported: two devices can both show a one per cent residual and need shot
counts a hundred apart.

\emph{(ii) At the protocol's budget the honest verdict is ``undecided.''} The floor is
$1.1\times10^{7}$ and $6.7\times10^{6}$ shots per component at the reference
point (a parametric bootstrap that frees the alternative lifts these to
$1.5\times10^{7}$ and $1.1\times10^{7}$; Sec.~\ref{sec:scope}), the median of
each direction over sampled working points is $8.9\times10^{8}$
and $1.1\times10^{7}$, respectively, and over sampled working points and directions it spans
$4.6\times10^{5}$--$2.1\times10^{11}$; at $10^{5}$
the residual measured is $5\%$, an order of magnitude above the bias, and the
power equals the level. An acceptance threshold that carries no sampling
distribution fails for the same reason,
the familiar ``residual below one per cent'' included: it rejects the correct
family everywhere.

\emph{(iii) That price belongs to the design, not to the device.} A residual is
usually reported as the device's, but the design decides what the model has to
explain: it fixes which syndromes are corrected and where, so it fixes both
fingerprints. Rebuilding them under each design and refitting moves $\rhob$ by
$8.2$ on the dispersive platform and $1.3$ on the exchange one, and the floor by
$68$ and $1.7$; the full recovery the reference floor is priced at is itself
dominated by the design that sets the cheap end of that range. The two objectives then disagree---the most precise design on the
dispersive platform costs $1.47\times10^{7}$ shots for the verdict against
$1.44\times10^{6}$ for a coarser one---and across the sixty-four designs precision
and detectability are nearly uncorrelated.

\emph{(iv) The design problem is about placement, not counting.} All sixty-four
placements are full rank in both families; what separates them is conditioning. At
three gates the local-angle rms varies by factors of $2.7$ (dispersive) and $3.9$
(exchange) with placement alone, while the optimum over counts is flat---four
gates at $41.41$~mrad against three at $41.72$ on the dispersive platform, a
$0.7\%$ margin inside a $2.7$-fold placement spread, and three at $32.21$ against
four at $35.48$ on the exchange one. The design selected
on the dispersive platform is the unique global optimum on the exchange one and
$0.7\%$ off the optimum on the dispersive one; optimising the couplings instead
puts it off both fronts.

\emph{(v) The family verdict depends on the design.} Eleven parameters against
thirteen forces a normalisation, and the normalisation interacts with the design:
four of seven conventions favour the larger family at the six-gate decoder, all
seven at the three-gate design. The mechanism is identifiable---the three-gate
design removes precisely the generators that the exchange family adds.

A companion note \cite{note}---cited as Ref.~\onlinecite{note} and called \emph{the
note} below---asks how many recovery gates a repetition-code decoder
needs, and answers it for three, five and seven qubits by exhaustive enumeration;
two of the results above, (iii) and (iv), bear directly on it. The two papers share
the fingerprint, the design algebra and the rank machinery, and differ in the
scale at which each asks its question: the note holds the budget high and the
objective wide---every parameter of the generator---so that the gate count is
the only variable, while this paper holds the objective narrow---the nine
local angles the code is built to see---and the budget at the value a run
would actually spend, so that the design is priced as an experiment rather
than counted as inventory. Section~\ref{sec:note} aligns the two line by line,
reproduces the note's table to $0.3\%$, and shows that the remaining
differences are exactly that choice of scale.

Taken together these say that the two decisions form a single problem whose answer
is a function of three numbers---the absolute bias $\rhob S$, the conditioning of
the design, and the shot budget---none of which is a property of the platform
alone. The first two are read off one Jacobian, the one the design builds, which
is why the price of a family verdict is itself a design variable.
Section~\ref{sec:protocol} turns this into a five-step protocol that can be run
before any beam time is booked.

The mechanism behind (iii) is worth naming before the details, because it is the
part of this paper that travels. Whether a design can move the price of a verdict
depends on which block of the fingerprint the misspecification reaches
(Sec.~\ref{sec:coupling}). A bias confined to the reference block, which every
design measures identically, is design-blind; a bias that reaches the code blocks
whose Jacobians the design reshapes is not, and here by a factor of sixty-eight.
The two platforms of this paper are the two ends of that axis, which is what turns
the scale of the effect from a limitation of three qubits into a rule.

\section{Setting}
\label{sec:setting}

\subsection{Fingerprint, families, and the residual}

The fingerprint is a stack of three single-logical-qubit Pauli transfer matrices:
the physical reference channel and the two first-level code channels of a
three-qubit repetition code \cite{main}, sixteen real components each, forty-eight
in all. For a parameter vector $\theta$ in a family $f$ the fingerprint is
$F_f(\theta)$; the construction, the code projectors and the design algebra are
those of Refs.~\cite{note,supp}.

The two families are
\begin{align}
\mathtt{zz}&:\ \mathfrak{su}(2)^{\otimes3}\oplus
 \mathrm{span}\{Z_{j}Z_{j+1}\}, & p&=11,\\
\mathtt{xy}&:\ \mathfrak{su}(2)^{\otimes3}\oplus
 \mathrm{span}\{X_{j}X_{j+1},\,Y_{j}Y_{j+1}\}, & p&=13 .
\end{align}
They are not nested. $X_jX_{j+1}$ and $Y_jY_{j+1}$ commute, so the exchange spans
a two-dimensional abelian plane, and $Z_jZ_{j+1}$ is orthogonal to that plane in
the Pauli inner product; a model drawn from one family therefore cannot represent
a device belonging to the other. Which family applies is fixed by the hardware:
fixed-frequency transmons have the dispersive term always on, tunable couplers
have the exchange.

For a fit of family $f$ to a device we use the normalised residual
\begin{equation}
\rho \;=\; \frac{\lVert F_{f}(\hat\theta)-F_{\mathrm{dev}}\rVert}{S},
\qquad
S \;=\; \lVert F_{\mathrm{dev}}-F(0)\rVert ,
\label{eq:rho}
\end{equation}
with $F(0)$ the fingerprint at zero coupling, so $\rho$ is the fraction of the
device's own distance from that limit which the family cannot represent.
$\rhob$ is the value $\hat\rho$ converges to---the residual of a noiseless
fit---while $\hat\rho$ itself is what a run reports, with standard error
$\sqrt{\nu}\sigma/S$ (Sec.~\ref{sec:stat}). What fixes
detectability is the product $\rhob S$---a difference of
Pauli-transfer-matrix components, dimensionless but neither a ratio of fidelities
nor a distance in Hilbert space---and it is this difference, not the
fraction $\rhob$, that enters the floor squared. The fit
is a multi-start Levenberg--Marquardt least-squares fit of \eqref{eq:rho}, always
started from the device parameters rewritten in the target family's layout. It is
deliberately generous: whatever residual survives is the family's, not the
optimiser's, and adding three perturbed starts changes it by less than one part in
$10^{4}$ at the four points where we checked.

The two archetypes are compared at equal physical strength: the exchange
couplings are scaled by $0.7069$ so that the per-layer coherent infidelity
$1-|\mathrm{Tr}\,U/d|^{2}$ \cite{nielsen} equals $1.295\times10^{-2}$ on both. At
the reference point this gives $S=0.3729$ for the dispersive device and $S=0.3378$
for the exchange one.

\subsection{The adequacy statistic}
\label{sec:stat}

With $N$ shots per fingerprint component the entry noise is $\sigma=1/\sqrt{N}$.
The residual of \eqref{eq:rho} is then estimated with standard error
$\sigma_\rho=\sqrt{\nu}\,\sigma/S$, where $\nu=48-p$ is the degrees of freedom of
the family being \emph{fitted}, so the natural adequacy statistic is
\begin{equation}
T \;=\; \Bigl(\frac{S\rho}{\sigma}\Bigr)^{2} \;=\; \rho^{2}S^{2}N ,
\label{eq:T}
\end{equation}
which is $\chi^{2}_{\nu}$ when the family is correct and non-central
$\chi^{2}_{\nu}(\lambda)$ with $\lambda=\rhob^{2}S^{2}N$ when it is wrong by a
bias $\rhob$. Everything in Sec.~\ref{sec:detect} follows from \eqref{eq:T}; no
Monte Carlo is involved, and the criterion becomes a formula that can be
evaluated before the experiment is scheduled.

Table~\ref{tab:cells} evaluates the two mismatch cells. The residual of the wrong
family is $0.42\%$ and $0.59\%$ of the device's own signal, and the corresponding
local angles move by $47\%$ and by $25\%$: the wrong family absorbs an
unrepresentable coupling into rotations, which is a bias and not a variance.

\begin{table*}[t]
\caption{\label{tab:cells}The two mismatch cells at the reference point.
$\rhob$ and $S$ as in \eqref{eq:rho}, $\nu=48-p$ the degrees of freedom of the
fitted family, $\sqrt{\nu}\sigma/S$ the standard error of $\hat\rho$ at
$N=10^{5}$. Both $\rhob$ and $S$ are in Pauli-transfer-matrix units, so $\rhob S$
is a difference of fingerprint components---dimensionless, but not a ratio of
fidelities. The shot floor $\Nstar$ is \eqref{eq:Nstar}; power is evaluated at
the protocol budget $N=10^{5}$. Both cells are at the full six-gate
recovery---\texttt{build\_K} in the notation of Refs.~\cite{note,supp}, the
design that corrects every site of both codes; Sec.~\ref{sec:coupling} re-prices
them under the other sixty-three.}
\small
\setlength{\tabcolsep}{9pt}
\begin{tabular}{lccccccc}
\toprule
cell & $\rhob$ & $S$ & $\nu$ & $\sqrt{\nu}\sigma/S$ & power & $\Nstar_{.05,.8}$ & $\Nstar_{.05,.95}$\\
\midrule
ZZ device, XY fit & $4.20\times10^{-3}$ & $0.3729$ & 35 & $5.02\times10^{-2}$ & $0.053$ & $1.07\times10^{7}$ & $1.54\times10^{7}$\\
XY device, ZZ fit & $5.90\times10^{-3}$ & $0.3378$ & 37 & $5.69\times10^{-2}$ & $0.056$ & $6.72\times10^{6}$ & $9.66\times10^{6}$\\
\bottomrule
\end{tabular}
\end{table*}

\section{First decision: can the family be decided?}
\label{sec:detect}

\subsection{The shot floor}

Fixing the level $\alpha$ fixes the threshold $c=\chi^{2}_{\nu}(1-\alpha)$; the
power is $1-F_{\chi'^{2}}(c;\nu,\lambda)$, monotone in $N$. Solving for the budget
gives
\begin{equation}
\boxed{\;\Nstar \;=\; \frac{\lambda_{\mathrm{req}}(\alpha,1-\beta,\nu)}
{(\rhob S)^{2}}\;}
\label{eq:Nstar}
\end{equation}
with $\lambda_{\mathrm{req}}$ the non-centrality whose tail beyond $c$ equals
$1-\beta$. The derivation presumes entry noise $\sigma=1/\sqrt{N}$ and nothing
else; Sec.~\ref{sec:scope}(iv) adds a systematic term $\sigma_{\mathrm{s}}$ on a
fingerprint component, which multiplies the floor by $1/(1-\Nstar/N_{c})$---where
$N_{c}\equiv\sigma_{\mathrm{s}}^{-2}$ is the shot budget at which that systematic
equals the shot noise---and can remove it altogether. For $\alpha=0.05$ and $80\%$ power one finds
$\lambda_{\mathrm{req}}=26.1$ at $\nu=35$ and $26.7$ at $\nu=37$---a $2\%$ change
across the two families, consistent with the $\sqrt{\nu}$ scaling that
\eqref{eq:Nstar} inherits from the $\chi^{2}$ tail. At $\alpha=0.01$ the constant
rises to $34.7$ and $35.5$. The floor is therefore controlled by one combination,
the effect size $\rhob S$, and by almost nothing else.
Sample costs of this $1/\text{effect}^{2}$ form are generic to estimation
\cite{haah}; what is special here is that the effect is a distance between two
models, which no target state defines. The logic is the one behind
confidence intervals in randomized benchmarking \cite{wallman}, with a model
comparison in place of a fidelity as the target.
The statistic is textbook---a non-central $\chi^{2}$ against a fixed alternative,
with power monotone in the non-centrality---so
$\lambda_{\mathrm{req}}$ is read from the standard tables \cite{kay}; what is
not is that the alternative is a change of model, not a shift of a parameter.

Table~\ref{tab:cells} evaluates \eqref{eq:Nstar}: $1.1\times10^{7}$ and
$6.7\times10^{6}$ shots per component, or of order $5\times10^{8}$ events across
the forty-eight components, against the $10^{5}$ per component that the protocol
assumes---two orders of magnitude more instrument time, at any readout rate, than
the runs whose residuals the protocol quotes. The standard-error column is the
reason. At $N=10^{5}$ the standard
error of $\rhob$ is $5.0\times10^{-2}$, twelve times the bias it is meant to
reveal, so the residual that would be reported is the noise and not the model
error. The power at that budget, $0.053$, is indistinguishable from the level:
the budget, not the statistic, is what fails, and the honest output of the
protocol there is the ``undecided'' of Sec.~\ref{sec:protocol}, not a family.
The same formula prices the remedy: raising the budget to $10^{6}$ lifts the
power to $0.09$ and $0.12$ in the two directions, $10^{7}$ reaches $0.77$ and
$0.96$, and the floor of Table~\ref{tab:cells} is where the power crosses $0.8$.

This invalidates any fixed threshold, and the one most often written down is no
exception. We do not attribute the rule ``accept the family if the residual is
below one per cent'' to a named protocol---it is quoted here because it is the shape
a criterion takes when the sampling distribution is left out---and the defect is
not in the choice of one per cent. An acceptance rule $\hat\rho\le t$ is a rule with
a domain. At $10^{5}$ shots a correct family reports $\hat\rho\simeq5\%$, so
$t=1\%$ rejects the correct family at essentially every point. The number that
$\hat\rho$ should be compared against is not
zero but $\sqrt{\chi^{2}_{\nu}(1-\alpha)}\,\sigma/S$, which is what \eqref{eq:T}
says and what Table~\ref{tab:cells} tabulates. A residual quoted without its shot
budget is not a measurement of the model. Error bars that are proved from a finite
sample rather than read off an asymptotic covariance are what makes such a
statement checkable \cite{sugiyama}; the floor above is that idea applied to a
difference between two models instead of a distance from a target.

It is worth writing the window out, because it shows what a fixed threshold would
have to be.
At level $\alpha$ a family is accepted when
$\hat\rho\le\rho_c(N)=\sqrt{\chi^{2}_{\nu}(1-\alpha)/(S^{2}N)}$, which at
$\nu=35$, $\alpha=0.05$ and the signal used here gives $19\%$ at $10^{4}$ shots,
$6.0\%$ at $10^{5}$, $1.9\%$ at $10^{6}$ and $0.60\%$ at $10^{7}$. A one-per-cent
window at $10^{5}$ shots would need $S=2.2$, six times the signal of this device;
at its actual signal the window reaches one per cent only at $3.6\times10^{6}$
shots. Below that the rule is not conservative---it is unsatisfiable.

\emph{Relation to non-nested model selection.} The two families are separate
hypotheses, not a nested pair, and the literature on such comparisons is
old and classical \cite{cox}: tests of separate families, and likelihood-ratio
selection in the general form of Vuong \cite{vuong}. What Vuong's statistic
needs is the sample size at which the expected log-likelihood difference of the
two fitted families exceeds its own fluctuation, the reciprocal square of their
residual difference---the $7.1\times10^{7}$ shots per component of
Sec.~\ref{sec:scope}(iii) is exactly that number. Penalty criteria sit in the
same accounting: an AIC or BIC comparison of an eleven- against a
thirteen-parameter family carries a fixed penalty that at $N\ge10^{5}$ is
negligible, so the criterion degenerates to a residual comparison, and the
sample at which the penalty binds again is set by the same $(\rhob S)^{-2}$ as
\eqref{eq:Nstar}. The upper-tail goodness-of-fit test used here is preferred
because it prices each family separately and so survives the case, likely in
practice, that neither family is true; the non-nested comparison is its
complement, not its replacement.

\subsection{Operating characteristics}
\label{sec:roc}

\begin{figure*}[t]
\centering
\includegraphics[width=0.86\textwidth]{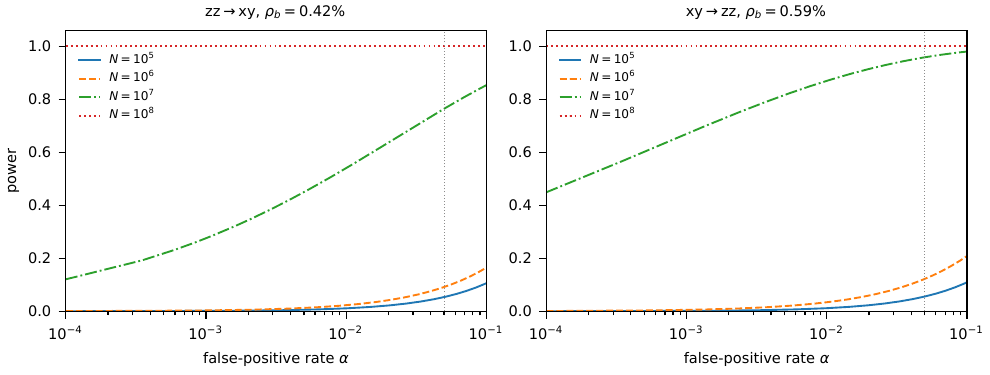}
\caption{\label{fig:roc}Receiver-operating characteristic of the family test at
$N=10^{5}$, $10^{6}$, $10^{7}$ and $10^{8}$ shots per component, for the two
mismatch directions. The non-centrality is $\lambda=\rhob^{2}S^{2}N$, so the
curves move left as $N$ grows. The dotted vertical line is $\alpha=0.05$. At the
protocol budget the curve is the diagonal.}
\end{figure*}

Figure~\ref{fig:roc} shows the receiver-operating characteristic of the family
test at four budgets. Since $\lambda=\rhob^{2}S^{2}N$, each decade of $N$
multiplies the non-centrality by ten, and the ordering of the two directions is
set by $\rhob S$, not by $\rhob$. At $10^{5}$ the curve lies on the diagonal, which is
the graphical form of the statement that the power equals the level; at $10^{7}$
it has separated in both directions.

\begin{figure}[t]
\centering
\includegraphics[width=\columnwidth]{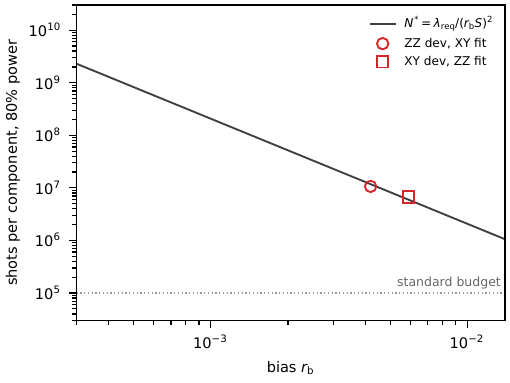}
\caption{\label{fig:shots}The shot floor \eqref{eq:Nstar} against the bias, for
$80\%$ power at $\alpha=0.05$ and $S=0.355$ (the mean of the two cells).
Symbols are the two cells of Table~\ref{tab:cells}; drawing the curve at the
mean $S$ displaces them along the axis by $+10\%$ and $-9\%$ in $\Nstar$, well
inside the decade scales drawn. The dashed line is the
protocol budget $10^{5}$: both cells sit a hundred times above it.}
\end{figure}

Figure~\ref{fig:shots} plots the floor against the bias. Because
$\Nstar\propto(\rhob S)^{-2}$, a factor of ten in the bias is two decades in the
budget. The two cells fall within a factor $1.6$ of each other even though their
fractional residuals differ by $1.4$, because $S$ compensates; and the horizontal
line marking $10^{5}$ is crossed nowhere in the plotted range, since even the
strongest bias drawn there still costs $10^{6}$.

\subsection{Across working points}
\label{sec:scan}

The bias is not a constant of the platform. It is the distance from the device
point to the model family, and both move with the point. Over sixteen sampled
working points (local rotation vectors of norm drawn uniformly in $5$--$50$~mrad,
couplings uniform in $\pm50$~mrad, the exchange scaled to the same per-layer
infidelity; the sample is drawn from one seeded generator, seed $20261108$, fixed
by the deposited \texttt{nm\_01}, and the forty-point sample of
Sec.~\ref{sec:points} begins with these sixteen) the floor
spans $5.7$ orders of magnitude, from $4.6\times10^{5}$ on the exchange side to
$2.1\times10^{11}$ on the dispersive one per
component (Table~\ref{tab:scan}). Even the cheapest of the thirty-two sampled
cells needs $4.6$ times the protocol budget.

\begin{table*}[t]
\caption{\label{tab:scan}The two mismatch cells over sixteen sampled working
points: the relative residual, the absolute residual $\rhob S$ in Pauli-transfer
units, and the shot floor of \eqref{eq:Nstar} at $80\%$ power and $\alpha=0.05$.
The last row is the reference point of Table~\ref{tab:cells}: the easy corner in
the dispersive direction, and mid-range rather than extreme in the exchange one.}
\small
\setlength{\tabcolsep}{3.5pt}
\begin{tabular}{lcccc}
\toprule
 & \multicolumn{2}{c}{ZZ device, XY fit} & \multicolumn{2}{c}{XY device, ZZ fit}\\
\cmidrule(lr){2-3}\cmidrule(lr){4-5}
 & range & median & range & median\\
\midrule
$\rhob$ & $8.9\times10^{-5}$--$1.2\times10^{-2}$ & $2.3\times10^{-3}$ &
$1.6\times10^{-4}$--$1.7\times10^{-1}$ & $2.8\times10^{-2}$\\
$\rhob S$ & $1.1\times10^{-5}$--$1.2\times10^{-3}$ & $1.7\times10^{-4}$ &
$2.3\times10^{-5}$--$7.6\times10^{-3}$ & $1.6\times10^{-3}$\\
$\Nstar$ & $1.8\times10^{7}$--$2.1\times10^{11}$ & $8.9\times10^{8}$ &
$4.6\times10^{5}$--$5.1\times10^{10}$ & $1.1\times10^{7}$\\
\midrule
reference point & \multicolumn{2}{c}{$4.20\times10^{-3}$, $\Nstar=1.07\times10^{7}$}
 & \multicolumn{2}{c}{$5.90\times10^{-3}$, $\Nstar=6.72\times10^{6}$}\\
\bottomrule
\end{tabular}
\end{table*}

Two things follow. First, the requirement is a property of the working point in
the same sense that the precision is, so it has to be quoted with one; a protocol
that quotes a single number is quoting its best case. The reference point used
throughout is the easy corner of the dispersive direction, where the median floor
of that direction is $83$ times higher; in the exchange direction it is
not an extreme at all---its relative residual is $4.7$ times smaller than the
median, but the median point's own $S$ is six times smaller still, so the two
absolute biases differ by a factor of only $1.25$ and the two floors by $1.6$.
The numbers of Table~\ref{tab:cells} are therefore a lower bound on what the
decision costs, a strong one in the first direction and a weak one in the second.

Second, $\rhob$ is the wrong currency. $S$ ranges over $0.016$--$0.21$ across the
sample against $0.373$ at the reference point, a factor of thirteen, which is
comparable to the spread of $\rhob$ itself; the product $\rhob S$ is what enters
\eqref{eq:Nstar}. This is also why the quoted $0.42\%$ and $0.59\%$ read as small
and are not: they are fractions of unusually large signals. And the ordering of
the two directions is not stable across points, since the exchange device is the
easier one at the reference point ($6.7\times10^{6}$ against $1.1\times10^{7}$)
and the harder one at the median ($1.1\times10^{7}$ against $8.9\times10^{8}$).

\subsection{Scope of the criterion}
\label{sec:scope}

Four qualifications belong here rather than in a limitations paragraph, because
they change how the number is used.

(i) \emph{The families are treated as exhaustive.} If the device is in neither,
both fits are rejected and no budget produces an acceptance. That is the correct
behaviour and a useful diagnostic: a device that fits neither archetype is a
result, not a failure of the test.

(ii) \emph{The degrees of freedom are those of a linear model.} The fingerprint is
nonlinear in $\theta$, so $\nu=48-p$ is exact only asymptotically. At
$\nu\simeq35$ the error is small compared with the $\sqrt{\nu}$ dependence, and
the power at the budget is so far from one that no correction of this order can
change the conclusion. Three further premises are made here: the
forty-eight components are independent (each is a separate circuit setting), the
binomial entry noise is treated as Gaussian, and the non-centrality
$\lambda=\rhob^{2}S^{2}N$ freezes the alternative at its pseudo-true fit---the
$p$ parameters of the wrong family, refitted, absorb part of the bias and lower
the effective non-centrality by a fraction the closed form does not track. A
parametric bootstrap of $T$ at the reference point (\texttt{nm\_13} of the
deposit)---synthetic binomial counts at the true device parameters, refitted
under the wrong family through the deposited pipeline, $150$ replicates per
experiment---checks all three at once, and the premises hold in the direction
that matters. The size comes out at $0.013$ and $0.007$, below the level: the
test is conservative, so the budget-priced verdict above is, if anything, less
likely to fire by accident than
quoted. The power at the floor is where the frozen alternative bites: the
empirical rates are $0.63$ and $0.51$, with a Monte Carlo standard error of
$0.04$ at $150$ replicates, against the $0.80$ the closed form predicts. The
mean of $T$ over the same replicates says why---it falls short of
$\nu+\lambda$ by $29\%$ and $39\%$ of $\lambda$, the fraction of the
non-centrality the refit absorbs---and those absorbed fractions map through the
same non-central law to powers $0.60$ and $0.51$. The two readings of one run
therefore agree within a Monte Carlo standard error, and both say the same
thing: the floor is a lower bound on the price of an eighty-per-cent-power
verdict, and at the absorbed fractions it corrects to
$1.5\times10^{7}$ and $1.1\times10^{7}$ shots per component. A second point per
direction---the working point whose own floor is nearest the median of its
direction in Table~\ref{tab:scan}, $1.0\times10^{9}$ and $1.3\times10^{7}$
shots per component, some $14\%$ above the column median there, and the point
where the \texttt{nm\_14} of the deposit repeats the experiment---gives less absorption
than the reference point, not more: power $0.67$ in both directions (standard
error $0.04$) against
$0.80$, sizes $0.033$ and $0.007$, and $23\%$ and $21\%$ of the
non-centrality absorbed, which corrects the local floor to
$1.3\times10^{9}$ and $1.6\times10^{7}$. The reference corner is the harder of
the two, so the correction above is the conservative one.

(iii) \emph{The per-component noise is $1/\sqrt{N}$, without readout penalty.} A
confusion-matrix correction at $99\%$ assignment fidelity costs $6$ to $8\%$ in
shots---$6\%$ for a symmetric model, $7.7\%$ for the register measured in
Sec.~\ref{sec:anchor}, where inverting the confusion matrix inflates the variance
of a component by $\prod_i f_i^{-2}$ over the three fidelities---and moves
nothing else. Telling the two
directions apart rather than detecting either is dearer: the residuals differ by
$\Delta\rho=1.70\times10^{-3}$, and resolving that difference costs
$7.1\times10^{7}$ shots per component, seven to eleven times the floor for
detecting either.

(iv) \emph{The residual is coherent.} Equation~\eqref{eq:Nstar} counts shot noise
and nothing else. A systematic uncertainty $\sigma_{\mathrm{s}}$ on a fingerprint
component enters the same statistic in quadrature,
$T=(\rho S)^{2}/(N^{-1}+\sigma_{\mathrm{s}}^{2})$, so the floor it implies is
\begin{equation}
\Nstar=\frac{\lambda_{\mathrm{req}}}{(\rhob S)^{2}-\lambda_{\mathrm{req}}\sigma_{\mathrm{s}}^{2}}
=\frac{\Nstar_{0}}{1-\Nstar_{0}/N_{\mathrm{c}}},
\label{eq:syst}
\end{equation}
with $N_{\mathrm{c}}\equiv\sigma_{\mathrm{s}}^{-2}$ the budget at which the
systematic equals the shot noise---$\sigma_{\mathrm{s}}$ is per component, in the
same units as $N^{-1/2}$, so $N_{\mathrm{c}}$ is a shot count and no $S$
enters---and $\Nstar_{0}$ the shot-only floor. While
$\Nstar_{0}\ll N_{\mathrm{c}}$ the
correction is a few per cent and can be absorbed; once $\Nstar_{0}\ge N_{\mathrm{c}}$
the floor \emph{does not exist}, because past that budget the residual measures the
systematic rather than the family, and no number of shots recovers the test. This
is the quantitative form of the regime boundary, and it is not specific to coherent
noise: any unmodelled term that moves the fingerprint without averaging away does
the same thing. Whether a term caps the floor is decided by shape rather than by
size. One that rescales the generator uniformly keeps the families nested and is
absorbed by the fit, at a finite multiplicative cost; one that shifts it is the
$\sigma_{\mathrm{s}}$ above. On the register of Sec.~\ref{sec:anchor} the shift is
the uncorrected readout bias, $\sigma_{\mathrm{s}}=1.67\times10^{-2}$, giving
$N_{\mathrm{c}}=3.6\times10^{3}$, while the gate error there is the rescaling: a
depolarising layer that costs $1.53$ on the floor and never makes \eqref{eq:syst}
diverge. We take the whole uncorrected bias as $\sigma_{\mathrm{s}}$, since nothing
in the run bounds the removable part; a tenfold smaller $\sigma_{\mathrm{s}}$
would still leave $N_{\mathrm{c}}$ a factor of thirty below $\Nstar_{0}$.
Section~\ref{sec:anchor} evaluates \eqref{eq:syst} on a measured
register, where the denominator is negative.

\section{Second decision: which recovery design?}
\label{sec:design}

\subsection{The design space}

The recovery step applies a Pauli correction at a subset of the three sites
before the next parity check, separately for the bit-flip and the phase-flip
syndrome. Each code therefore chooses one of eight subsets, and a design is a pair
of subsets, sixty-four in all. The gate count is the sum of the two cardinalities,
so the multiplicities are those of the sum of two binomials,
\begin{equation}
1,\ 6,\ 15,\ 20,\ 15,\ 6,\ 1 \quad\text{for } G=0,\dots,6 ,
\label{eq:binom}
\end{equation}
and every design in both families is full rank. We verified this by exact
rank arithmetic on the Jacobian of the fingerprint: no design of the sixty-four is
rank deficient in either family, so at this register size the design problem
contains no feasibility constraint at all. What separates the designs is
conditioning, and conditioning alone.
A design enters an estimation problem only through its information matrix, and
the map from design to matrix is the subject of the classical theory of
optimal design \cite{pukelsheim}; the sixty-four placements are a finite
instance of that theory and the rms over parameters is one of its criteria.

The objective is the Cram\'er--Rao rms of the nine local angles at $N=10^{5}$ per
component, computed from the inverse Fisher information of the design's Jacobian.
We use the local block because it is the block both families share; the couplings
are reported separately where they matter.
Targeting a few named parameters rather than a full gate set is the practice
that robust phase estimation established \cite{kimmel}; the nine angles are
that kind of target, and the point is that even so small a target has a
design-dependent price.

\subsection{The budget curve}

\begin{figure*}[t]
\centering
\includegraphics[width=0.86\textwidth]{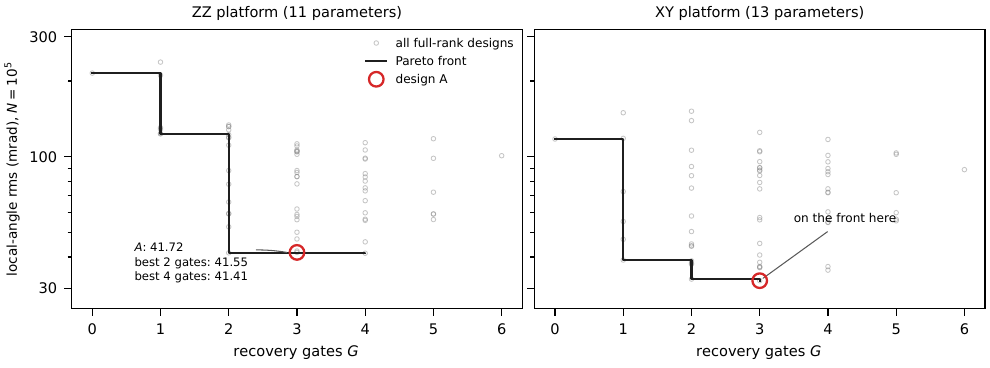}
\caption{\label{fig:pareto}Every full-rank design (grey circles), the Pareto front
in (gates, local-angle rms) (black), and design $A$ (red), for the two platforms
at their own device point and $N=10^{5}$ per component. On the exchange platform
$A$ is the unique global optimum of the sixty-four designs in this objective. On
the dispersive one it is not on the front: two gates reach $41.55$ and four reach
$41.41$ against $A$'s $41.72$. Two designs are worth locating by their numbers
rather than by their position: the cheapest to test, $\Nstar=1.4\times10^{6}$, and
the most precise, $\Nstar=1.5\times10^{7}$, lie a decade apart on the vertical axis
while differing by a few per cent on the horizontal one---which is the correlation
this figure exists to show.}
\end{figure*}

Table~\ref{tab:budget} is the best full-rank rms at each gate count, on each
platform at its own device point, and Fig.~\ref{fig:pareto} shows the same data as
a Pareto front in the (gates, rms) plane.

\begin{table}[t]
\caption{\label{tab:budget}Best local-angle rms (mrad) at each gate count, on each
platform at its own device point, $N=10^{5}$ per component. The second column is
the number of designs at that count, \eqref{eq:binom}. Design $A$ is
bit~$\{1\}$, phase~$\{1,2\}$.}
\footnotesize
\setlength{\tabcolsep}{3pt}
\begin{tabular}{lcccc}
\toprule
$G$ & designs & ZZ (11 par.) & XY (13 par.) & note\\
\midrule
0 & 1  & 215.0 & 117.7 & no correction\\
1 & 6  & 123.4 & 38.9  & \\
2 & 15 & 41.55 & 32.68 & best two-gate\\
3 & 20 & 41.72 & 32.21 & design $A$\\
4 & 15 & \textbf{41.41} & 35.48 & best overall, ZZ\\
5 & 6  & 56.34 & 55.69 & \\
6 & 1  & 101.0 & 88.9  & six-gate decoder\\
\bottomrule
\end{tabular}
\end{table}

The two platforms answer differently. In the local-angle objective the exchange
platform makes $A$ the unique global optimum of all sixty-four designs, at
$32.21$~mrad, better than the best two-gate design by $1.5\%$ and better than
the
six-gate decoder by a factor of $2.76$. On the dispersive platform $A$ is not on the
front at all: the two-gate design that corrects the phase code at sites $\{1,2\}$
and leaves the bit code alone reaches $41.55$ with one gate fewer, and a
four-gate design reaches $41.41$. $A$ sits $0.4\%$ above the first and $0.7\%$
above the second.

The differences are small, but they are deterministic and they have a reading: on
a dispersive device the bit-code correction moves information out of the
fingerprint, and buying the third or fourth gate recovers at most $0.34\%$ over
the best two-gate design. The front on that platform is flat between two and four
gates---$41.41$ to $41.72$, a spread narrower than the difference between two
placements at the same count.

\subsection{Placement dominates cardinality}

That last remark is the quantitative core of the design problem, and it can be
stated as two ranges. At a fixed count of three gates the twenty designs span
$41.72$--$112.59$~mrad on the dispersive platform and $32.21$--$125.02$~mrad on
the exchange one: factors of $2.7$ and $3.9$. The best value over all gate counts
spans $41.41$--$215.0$ and $32.21$--$117.7$, factors of $5.2$ and $3.7$. Placement
and cardinality are therefore of comparable importance---on the dispersive
platform, choosing the right three sites is worth about as much as choosing the
right number of gates---so a report that quotes only the count has quoted half a
design.

\subsection{Does the placement survive a change of family?}

The count is not the whole design, so the same question has to be asked of the
pair: is the argmin pair a property of the family? Table~\ref{tab:bestpair}
answers it at the reference point, and the answer is mostly yes. Five of the
seven counts select the same pair on both platforms. The two that do not are
four gates---which happens to be the dispersive optimum---and five, and carrying
either choice across costs only $1.03$ and $1.02$ times as much, so even the
design that makes one platform optimal is within three per cent of optimal on
the other.

\begin{table}[!t]
\caption{\label{tab:bestpair}The argmin design at each gate count, on each
platform in its own family, with the local-angle rms in mrad at $N=10^{5}$ in
parentheses. $b\{.\}$ is the bit-code site set and $p\{.\}$ the phase-code one;
bold is the global optimum of that platform.}
\footnotesize
\setlength{\tabcolsep}{5pt}
\begin{tabular}{cll}
\toprule
$G$ & dispersive (ZZ) & exchange (XY)\\
\midrule
0 & --- (215.0)              & --- (117.7)\\
1 & $p\{2\}$ (123.4)         & $p\{2\}$ (38.9)\\
2 & $p\{1,2\}$ (41.55)       & $p\{1,2\}$ (32.68)\\
3 & $b\{1\}p\{1,2\}$ (41.72) & $b\{1\}p\{1,2\}$ (32.21)\\
4 & $b\{1,3\}p\{1,2\}$ (\textbf{41.41}) & $b\{1,2\}p\{1,2\}$ (35.48)\\
5 & $b\{1,2\}p\{1,2,3\}$ (56.34) & $b\{1,2,3\}p\{1,2\}$ (55.69)\\
6 & full (101.0) & full (88.9)\\
\bottomrule
\end{tabular}
\end{table}

Over the forty sampled points the agreement is weaker and uneven: $70\%$ at
three gates and $78\%$ at four, but $35\%$ at two and $22\%$ at one, with the
optimal count itself agreeing at half the points. Read as a statement about
argmins that is a negative result. Read as a statement about the surface it is
the expected consequence of the flatness of Sec.~\ref{sec:design}: carrying one
platform's pair to the other costs a median factor of $1.05$ at two gates and
$1.00$ at three, four and five, with ninetieth percentiles of $1.38$, $1.15$,
$1.31$ and $1.10$. Only the single-gate design is brittle, at $1.48$ median and
$18$ times in the worst case, because one gate has no redundancy with which to
absorb a wrong site.

The honest form of the statement is therefore not that the placement is
family-independent, which it is not, but that above two gates the argmin is not
a well-defined object on a surface this flat, while the cost of choosing wrongly
is. That is a stronger reason for the reporting rule of Sec.~\ref{sec:protocol}
than any single working point could give.

\subsection{Across working points}
\label{sec:points}

Forty sampled working points---the same seeded scheme as
Sec.~\ref{sec:scan}, extended to forty; this scan needs only Jacobians, no fits,
so it can afford the full sample---give the distribution of the optimal count, and it is not
concentrated. On the dispersive platform the argmin is four gates at $26$ points,
three at $13$ and two at $1$; on the exchange platform three at $23$, four at
$15$ and two at $2$. Design $A$ is the best of all sixty-four designs at four
of forty points in the first family and two of forty in the second. Three gates
beat two at $39$ and $38$ of forty points, with median ratios $0.866$ and $0.942$,
so the third gate usually pays---but the fourth often pays more, and neither
statement is a theorem about the platform.

The practical form of this is that the count and the placement have to be chosen
at the working point, from the same Jacobian that will be used in the fit. The
enumeration costs sixty-four small linear-algebra problems and no data.

\subsection{Mechanism}

Why the design matters so much is a conditioning statement, and it can be
localised. The six-gate decoder has condition number $396$ on the dispersive
platform and $324$ on the exchange one; the three-gate design has $198$ and $143$.
Removing the three-site correction lowers the local-angle rms by $2.42\times$ on
the first and $2.76\times$ on the second, so the design helps the
thirteen-parameter family more. The reason is that the correction the standard
decoder applies erases, from the fingerprint, precisely the operators that the
exchange family adds: the parameter directions that the larger family needs are
the ones that the full correction removes.

\subsection{Relation to the gate-count note}
\label{sec:note}

The note \cite{note} fixes the same sixty-four designs at three qubits and
recommends a different one---three gates, not four. Both recommendations are
right, and the comparison is worth a paragraph because a reader who sets the two
abstracts side by side will see a contradiction that is not there.

Both papers compute the same object, and this one is a companion to that one.
The fingerprint, the sixty-four designs, the gate count as a popcount and the exact
ranks are common to the two; what differs is the scale at which the question is
posed, and Table~\ref{tab:scales} sets the choices side by side. The note holds the
budget high and the objective wide so that the gate count is the only variable;
this paper holds the objective at the nine angles the code is built to see and the
budget at the value a run would spend, so that the design is priced as an
experiment. Neither placement is a correction of the other: a paper about which
design to build should assume that the verdict is affordable, and a paper about
whether to report a verdict should not.

\begin{table}[t]
\caption{\label{tab:scales}The two scale choices. The fingerprint, the design
space and the exact ranks are common; every other entry is a decision about what
to hold fixed. The last two rows give the question each paper asks and the form
of the answer it can give: on the left a count, three gates and $22\times$ over
six; on the right a price, $2.4\times$ for three gates against six in this metric
and $68\times$ across the designs.}
\begin{ruledtabular}
\begin{tabular}{lll}
 & note \cite{note} & this paper\\
\hline
Registers & $3,5,7$ & $3$; $5$ sampled\\
Objective & all $4n-1$ & nine angles\\
Noise model & gates, readout & coherent only\\
Budget & $10^{7}$ & $10^{5}$\\
Designs & all, at each $n$ & all at $n=3$\\
Question & the count & the price\\
Answer & $3$ gates, $22\times$ & $2.4\times$, $68\times$\\
\end{tabular}
\end{ruledtabular}
\end{table}

Every number the note quotes for this register is reproduced here. Its precision
table gives $6.36$, $5.57\times10^{-2}$, $2.62\times10^{-2}$, $1.68\times10^{-2}$,
$1.68\times10^{-2}$, $9.95\times10^{-2}$ and $3.75\times10^{-1}$~rad at
$0$--$6$ gates; recomputing them with the note's own definition---all $4n-1$
parameters, its noise table, $10^{7}$ shots---returns each of the seven to better
than $0.3\%$, three gates against six at $22.4$ where it quotes $22$, and three
against four at $1.00001$ where it says the two agree to four digits. The
minimum full-rank count is $0$ there, and the zero-gate design is full rank here,
$11/11$ and $13/13$. Its Lemma~1---the count depends on the site set only through
its cardinality---is, in this language, the statement that a design's count is
the popcount of its mask, which we verify for all sixty-four.

The two apparent disagreements are then bookkeeping. \emph{Three gates or four.}
On the dispersive platform four gates has the lower rms by $0.7\%$ in the
local-angle objective and by $1.8\%$ in the all-parameter one, and at the note's
own point the two agree to four digits; the placement spread at a fixed count is
$2.7$-fold. A margin that small inside a spread that large is a tie, broken
towards the cheaper design there and towards the argmin here. \emph{$2.4$ or
$22$.} The factor by which three gates beats the standard decoder belongs to the
working point, not to the design: the note measures $1.8$--$88$ over its forty
points and quotes $22$ at its own, and this paper's point returns $2.37$ in the
note's metric and $2.42$ in ours. Run at the note's point, the machinery here
returns $22.4$ and the note's optimal split, so what separates the two papers is
the point---about which both say the same thing.

\emph{Why those two budgets.} The note's $10^{7}$ is not a round number taken on
comfort. At its own point the design it recommends has $\Nstar=8.6\times10^{6}$,
so $10^{7}$ sits $1.2$ times above the floor of the recommended design and $2.7$
times below that of the six-gate decoder it argues against---its budget is placed
so that the count question is decidable and the verdict question does not arise.
The $10^{5}$ used here is what a run of the kind that quotes residuals spends,
taken as the representative low budget of the literature rather than as a
requirement; it lies two orders below both,
which is what makes the verdict, not the count, the binding constraint, and the
power interpolation of Sec.~\ref{sec:detect} prices what raising it buys. Read that
way, the two abstracts ask different questions with the same machinery, and the
answer to each is the answer to its own question.

\emph{Why those two objectives.} The rms over all $4n-1$ parameters answers
``which design determines the device''; the nine local angles answer ``which
design determines the thing the code is for''. The nine are a subspace of the
eleven, so the two are not independent measurements of the same quantity: at the
note's point the all-parameter rms spans $3.6$ over the sixty-four designs while
the verdict price spans $13$ ($2.10\times10^{6}$--$2.74\times10^{7}$ shots), and
the two orderings are uncorrelated (Spearman $-0.085$). That is the quantitative form of the statement that precision and
detectability are different functionals of the same Jacobian, and it is the reason
a single scale cannot carry both conclusions.

The one-line summary is that the note proves that cost and placement separate,
and this paper shows that the information side does not. The note's separation
stands: the design space can be enumerated once, and its Lemma~1 is what makes
the sixty-four problems tractable. What fails is the assumption that the
placement can be optimised once for all purposes, since it drives the error bars
and the detection floor in opposite directions (Spearman $-0.16$). At the note's
own point the refinement changes no recommendation---the three-gate design is
still the precision optimum, as the note says---and adds a price it does not
compute: the same design needs $4.1$ times the shots of a two-gate one for the
family verdict.

\section{The two decisions are one decision}
\label{sec:coupling}

\subsection{The design sets the price of the verdict}

Section~\ref{sec:detect} quotes one floor per platform, and both were computed at
one design: the full recovery, the six-gate one that \texttt{build\_K} implements.
The design is a choice. It fixes which syndromes are corrected and on which sites,
so it fixes the device's fingerprint and the model family's alike, and the residual
between them is a function of it.

We rebuild both under each of the sixty-four designs and refit. Two checks fix the
construction: at the full design the rebuilt fingerprint agrees with the published
one to $10^{-19}$, so the reduced designs are a strict generalisation and not a
different object; and the full-design row of Table~\ref{tab:designfloor} returns
the entries of Table~\ref{tab:cells} unchanged.

On the dispersive platform the residual is not a constant of the device. It runs
from $1.40\times10^{-3}$ to $1.14\times10^{-2}$, a factor $8.2$, and the floor with
it from $1.44\times10^{6}$ to $9.74\times10^{7}$, a factor $68$. On the exchange
platform the same sweep moves the residual by $1.3$ and the floor by $1.7$. A
verdict quoted without its design carries two digits it has not earned.

The two objectives do not agree, and the full design priced in
Table~\ref{tab:cells} does not survive the comparison. On the dispersive platform
the design with the
smallest error bars---two gates, correcting the phase code at sites $\{1,2\}$ and
leaving the bit code alone, the same design that beat $A$ in
Sec.~\ref{sec:design}---needs $1.47\times10^{7}$ shots for the verdict, ten times
the $1.44\times10^{6}$ of a four-gate design whose estimate is twice as coarse;
that four-gate design is also slightly more precise than the full recovery
($69.3$ against $70.3$, Table~\ref{tab:designfloor}), so the reference design is
strictly dominated at this point, and the factor $68$ above runs from the
cheapest design to the zero-gate one, a different design that is dominated on
both objectives. Every design keeps the floor above $10^{6}$, so the verdict is
far above any realistic budget whichever endpoint heads the range. The Pareto
front in the (rms, $\Nstar$) plane holds six designs on this platform and eight
on the exchange one, and it is steep at both ends: stepping off the most precise
design to rms $34.6$~mrad, within $1.2\%$ of the best, cuts the verdict price by
more than half, to $6.7\times10^{6}$; the next front point, rms $46.8$~mrad,
costs $6.3\times10^{6}$; and the cheapest verdict, $1.44\times10^{6}$, belongs
to an rms of $69.3$, a factor two off the best. Across the sixty-four designs
the two objectives are nearly uncorrelated (Spearman
$-0.16$ and $-0.09$): they are
different functionals of the same Jacobian, and the choice they force is a real
one---six designs out of sixty-four sit on the front, none at both extremes.

\begin{table}[!t]
\caption{\label{tab:designfloor}The family test re-priced by the design. Each row
rebuilds both fingerprints under that design and refits; $\rhob$, $S$ and $\Nstar$
are as in Table~\ref{tab:cells}, and rms is the local-angle Cram\'er--Rao rms of
the fitted family at $N=10^{5}$ under the same design. ``Full'' is the six-gate
recovery and reproduces Table~\ref{tab:cells} exactly. The rms column is that of
the \emph{fitted} family under the design, not the device's own-family rms of
Table~\ref{tab:budget}, and the two are not comparable. The label after each design
is its bit and phase mask, written in decimal over sites numbered from one---$3$
is $011$ and $6$ is $110$, so ``cheapest 3/6'' corrects sites $\{1,2\}$ of the
bit code and $\{2,3\}$ of the phase code, four gates in all;
``cheapest'' and ``most precise''
are the extremes of the two objectives over the sixty-four designs.}
\footnotesize
\setlength{\tabcolsep}{3.6pt}
\begin{tabular}{llcccc}
\toprule
 & design (bit/phase) & gates & rms & $\rhob$ & $\Nstar$\\
\midrule
ZZ & full 7/7 & 6 & 70.3 & $4.20\times10^{-3}$ & $1.07\times10^{7}$\\
   & cheapest 3/6 & 4 & 69.3 & $1.14\times10^{-2}$ & $1.44\times10^{6}$\\
   & most precise 0/3 & 2 & 34.2 & $3.63\times10^{-3}$ & $1.47\times10^{7}$\\
\midrule
XY & full 7/7 & 6 & 79.5 & $5.90\times10^{-3}$ & $6.72\times10^{6}$\\
   & cheapest 3/6 & 4 & 80.6 & $7.68\times10^{-3}$ & $3.97\times10^{6}$\\
   & most precise 0/3 & 2 & 36.5 & $6.06\times10^{-3}$ & $6.60\times10^{6}$\\
\bottomrule
\end{tabular}
\end{table}

\begin{figure*}[!t]
\centering
\includegraphics[width=\textwidth]{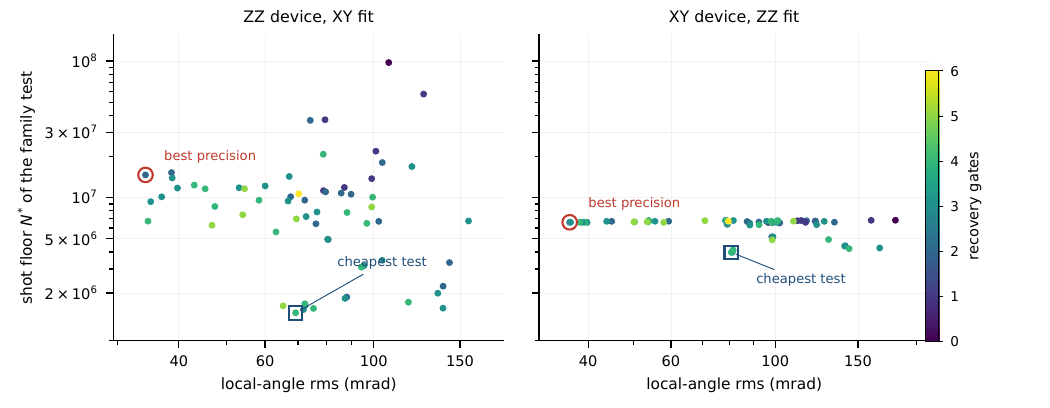}
\caption{\label{fig:designfloor}Every design as a point in the plane of the two
objectives, at the reference device point and $N=10^{5}$ per component: the
local-angle rms of the fitted family against the shot floor of the family test
that the same design implies. Colour is the gate count. The circled and squared
points are the most precise design and the cheapest-to-test design on each
platform; on the dispersive platform they are a factor ten apart in shots and a
factor two apart in rms.}
\end{figure*}

\subsection{The mechanism: which block the misspecification reaches}

The difference between the platforms has a mechanism, and it is the block structure
of the fingerprint. On the exchange platform $98.1\%$ of $\|\delta\|^{2}$ lies in
the reference block---the one block every design measures identically, since
$J_{\mathrm{ref}}$ enters every design---and the verdict is nearly design-blind. On
the dispersive platform only $48.0\%$ lies there, and $52.0\%$ lies in the two code
blocks whose Jacobians the design reshapes. \emph{A misspecification confined to
the single-qubit block cannot be made easier to detect by choosing a design; one
that reaches the code blocks can, and here by a factor of sixty-eight.}

That is a rule rather than a caveat. If the purpose of a run is to decide the
family, the design should be chosen for that decision and not for the error bars,
and the two differ enough to matter: a factor ten in shots. If the purpose is a
parameter estimate, the price of the verdict it also implies belongs in the report.

The rule also predicts, which is what separates it from a description of two
platforms. It says that a register whose code absorbs more of the discrepancy must
be more design-sensitive, and that is what the five-qubit run of
Sec.~\ref{sec:outlook} returns: there $71.8\%$ of the discrepancy sits in the code
block and the placement spread at a fixed count rises from $2.7$ to $119$. Two
readings of one mechanism at two register sizes are worth more here than a larger
sample at one.

\subsection{The averaging convention}

The families have eleven and thirteen parameters, so comparing them requires a
normalisation, and the normalisation is not neutral. We compute the
Cram\'er--Rao covariance of every parameter on each platform in its own family, at
two designs, and report seven conventions in Table~\ref{tab:norm}.

\begin{table*}[t]
\caption{\label{tab:norm}The platform comparison under seven normalisations, at
the six-gate decoder and at the three-gate design $A$. Entries are in mrad except
the last row of each block, which is the rms of $\sigma_i/|\theta_i|$ in units of
$10^{-3}$; the XY/ZZ ratio is dimensionless throughout, below one meaning the
exchange platform is better. The final line counts how many of the seven favour
XY.}
\footnotesize
\setlength{\tabcolsep}{4pt}
\begin{tabular}{lcccccc}
\toprule
 & \multicolumn{3}{c}{six-gate decoder} & \multicolumn{3}{c}{three-gate design $A$}\\
\cmidrule(lr){2-4}\cmidrule(lr){5-7}
normalisation & ZZ & XY & XY/ZZ & ZZ & XY & XY/ZZ\\
\midrule
per-parameter rms, local block & 101.0 & 88.9 & 0.880 & 41.7 & 32.2 & 0.772\\
per-parameter rms, couplings     & 61.1 & 62.9 & 1.029 & 38.5 & 25.8 & 0.669\\
per-bond coupling block          & 61.1 & 88.9 & 1.456 & 38.5 & 36.4 & 0.946\\
total Frobenius & 315.1 & 294.9 & 0.936 & 136.5 & 109.5 & 0.802\\
rms over all $p$ & 95.0 & 81.8 & 0.861 & 41.2 & 30.4 & 0.738\\
weakest direction & 234.2 & 197.4 & 0.843 & 117.2 & 86.8 & 0.740\\
rms of $\sigma_i/|\theta_i|$ ($10^{-3}$) & 4628 & 4687 & 1.013 & 2743 & 2058 & 0.750\\
\midrule
normalisations favouring XY & & & 4 of 7 & & & 7 of 7\\
\bottomrule
\end{tabular}
\end{table*}

At the six-gate decoder the verdict is not robust. The exchange platform is better
on the local angles by $12\%$ and on the weakest direction by $16\%$, but worse
per coupling by $3\%$, worse per bond by $46\%$, and worse relative to parameter
size by $1\%$. Four of seven conventions favour it, and the bond-level object---the
coupling error of a bond, which is what a device physicist would quote---is one of
the three that do not. The seven rows are also not seven independent verdicts:
the total Frobenius norm, the rms over all parameters and the per-parameter local
rms are dominated by the same nine local entries, so the count ``four of seven''
compresses to about three nearly independent dimensions---the local block, the
coupling block (per bond against per parameter) and the relative error. Read
that way the decoder verdict is: exchange better on the local dimension, worse
on both coupling dimensions.

At the three-gate design all seven favour the exchange platform, including the two
that reversed, and the two that reversed move far: the per-bond ratio goes from
$1.456$ to $0.946$ and the relative-error ratio from $1.013$ to $0.750$, so the
whole set now lies between $0.67$ and $0.95$. The two rows that flip are exactly
the two that weight the
couplings most heavily: per-bond quadrature and per-parameter relative error. That
is not an accident. The three-gate design improves conditioning on both platforms
but disproportionately on the one with more coupling directions, so the coupling
comparison, which is where the exchange platform was losing, is where it gains
most.

A sentence of the form ``the thirteen-parameter family is not worse'' therefore
has no truth value until a normalisation and a design are named. At the design a reader would use by
default it is true of the average over local angles and false of the bond-level
coupling; at design $A$---which is the exchange platform's optimum but not the
dispersive one---it is true of all seven, a comparison made at the same design on
both platforms and therefore free of the design-versus-family entanglement.
The correct sentence is narrower and stronger than either:
\emph{at the precision-optimal design the exchange archetype is better under every
normalisation we can defend; at the standard decoder its advantage is real on
every average except the per-bond coupling error, and which of the two leads
depends on the averaging convention.}

\subsection{The strength convention}
\label{sec:strength}

A third choice is made before either of these, and it is the one least often
written down: how the two archetypes are put at equal strength. We match the
per-layer coherent infidelity $1-|\mathrm{Tr}\,U/d|^{2}$, which fixes the
exchange couplings at $0.7069$ of the dispersive angles. The conventions in
common use collapse to two values of one scale. Average-gate infidelity, process
infidelity
and the diamond-norm distance to the identity are all monotone in the same
$1-|\mathrm{Tr}\,U/d|^{2}$ and reproduce $0.7069$; matching the generator's
Frobenius norm gives $1/\sqrt{2}$ analytically and $0.70686$ once the
non-commutation of the local and coupling layers is included, $0.01\%$ away;
matching the parameter angles themselves gives $1$. The freedom is a single
scale $s$ on $[0.71,1.00]$.

The verdicts of Secs.~\ref{sec:detect} and \ref{sec:design} survive that
interval in their ordering, and one of them strengthens with $s$. The design
conclusions are untouched in order---$A$ remains the unique global optimum on the exchange platform at every
$s$ tried, four gates remains the dispersive optimum---while their absolute
scale moves: the exchange rms runs from $36.2$ down to $23.2$~mrad across the
sweep, a factor $1.56$ end to end around the quoted $32.2$. The exchange
platform stays the easier one to test: the ratio $\Nstar_{\mathrm{xy}}/\Nstar_{\mathrm{zz}}$ runs from
$0.63$ at $s=0.7069$ to $0.16$ at $s=1$, and reversing it would take $s=0.63$,
an exchange coupling $11\%$ below the matched value---outside what any of the
conventions produces, though not by a wide margin. The mechanism sits in the two
factors. $S$ is constant to four digits across the sweep, $0.3378$ at $s=0.6$ and
$0.3379$ at $s=1.15$, because $99.997\%$ of $\|F_{\mathrm{dev}}-F(0)\|$ comes
from the nine local angles, which the sweep does not move; the entire effect
therefore travels through $\rho$, which grows from $4.3\times10^{-3}$ to
$1.6\times10^{-2}$ as the coupling the wrong family must absorb grows. Between
the two defensible conventions the floor moves by a factor $4.0$: less than the
factor $68$ that the design is worth, and five orders of magnitude less than the
working point. Of the three handles this is the weakest, which is the useful
thing to know about it.

What does move is the count of normalisations that favour the exchange platform,
and it moves in one direction. At the six-gate decoder four of the seven do so at
$s=0.7069$ and six at $s=1$; the objection is the same in both, the per-bond
coupling error, at $1.46$ and $1.20$. At the three-gate design all
seven favour it under either convention. The sentence that needs no
qualification is correspondingly the narrowest: at three gates the exchange
archetype is better under every normalisation and both conventions, while at the
standard decoder it is better on everything except the per-bond coupling error,
which no convention rescues.

\section{A protocol}
\label{sec:protocol}

The analysis collapses into five steps, all of which can be run before beam time
is booked.

\begin{enumerate}
\item Fix the setting table, the shot budget $N$ per component, and the candidate
families. Compute $\sigma=1/\sqrt{N}$ and, for the device point, $S$.
\item Fit every candidate family to the same data, from the same starting points,
and record the residual $\hat\rho_f$ of each.
\item Test adequacy family by family,
$\hat\rho_f^{2}S^{2}N\le\chi^{2}_{\nu_f}(1-\alpha)$. If exactly one family is
accepted, that is the answer. If both are accepted, $N$ is below the floor
\eqref{eq:Nstar} for that point: the honest report is that the data cannot choose. If neither is accepted, the device is in neither family, which
is a result and not a failure.
\item Enumerate the designs at the working point---sixty-four linear-algebra
problems---and report the count \emph{and} the placement of the optimum, together
with the spread over placements at that count. Above two gates the identity of
the optimum does not transfer between platforms while the cost of getting it
wrong does, so the spread is the quantity that carries over. When the family
decision is the purpose of the run, report the floor that each design implies as
well: at the
reference point here the most precise design and the cheapest-to-test design are a
factor ten apart, so the two cannot be chosen together by default.
\item When comparing families, state the normalisation, the design and the
strength convention. At the
standard decoder the ordering of the two archetypes depends on all three, and a
comparison that does not fix them is a comparison of conventions.
\end{enumerate}

Step 3 is the one that changes practice, because it says the decision has a price
and the price is two orders of magnitude above the one in use; an experiment that
reports a residual below one per cent at $10^{5}$ shots has reported its own noise.
Step 4 is the one that changes reporting, because it says the gate count is half of
the design---and, when the verdict is what the run is for, the other half is the
price of that verdict.

\begin{figure}[t]
\centering
\includegraphics[width=\columnwidth]{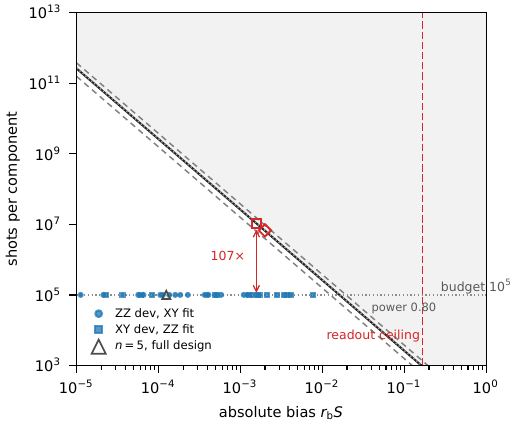}
\caption{\label{fig:phase}The two decisions on one plane. A configuration is
placed by the absolute bias $\rhob S$ it has to resolve and by the budget $N$ it
would spend; the heavy curve is power $0.8$,
$N=\lambda_{\mathrm{req}}/(\rhob S)^{2}$, the thin dashed curves are powers
$0.5$ and $0.95$, and the shading is the side on which the verdict is
affordable. Dots are the sixteen sampled working points of each cell, placed at
$10^{5}$; open symbols are the two reference cells, at their own floor, with the
arrow giving the factor to the budget; the triangle is the five-qubit full
design. The dotted curve is the same boundary for $\nu=37$, $2.3\%$ above the
one drawn. The red line is the tenth-margin line of Sec.~\ref{sec:anchor}: to
its left the register's uncorrected systematic exceeds a tenth of the signal
($\rhob S<1.7\times10^{-1}$). The existence boundary of \eqref{eq:syst}, where
the corrected floor ceases to be real, lies a further factor ten left, at
$\rhob S=8.5\times10^{-2}$; the plotted configurations lie to the left of both.}
\end{figure}

The three numbers of steps 1--3 are not independent, and Fig.~\ref{fig:phase} is
what their relation looks like. Write $x=\rhob S$ for the bias and $N$ for the
budget; the test rejects with power $1-\beta$ exactly when
$N\ge\lambda_{\mathrm{req}}(\alpha,1-\beta,\nu)/x^{2}$, so on log axes the
boundary is a line of slope $-2$, and the band between powers $0.5$ and $0.95$ is
a constant $2.4$ in $N$ at every bias, because the ratio of the two
non-centralities depends on $\alpha$ and $\nu$ alone. The dependence on the
hypothesis space is as weak: the boundary for $\nu=37$ lies $2.3\%$ above the one
for $\nu=35$, so a family added to the model set costs a hundredth of a decade of
budget, which is the quantitative form of the claim that widening the model set is
cheap next to moving the working point or the design.

What the plane adds is a place to stand. Every configuration computed here lies
below the curve, and the distances differ by five decades: the sixteen sampled
working points of each cell span $2.8$ decades of absolute bias, from
$1.1\times10^{-5}$ to $7.6\times10^{-3}$, the five-qubit full design sits at
$1.2\times10^{-4}$, and at $10^{5}$ shots each of them is short of the verdict by
a factor between five and $2\times10^{6}$. The two reference cells are short by
$107$ and $67$. The vertical line is the tenth-margin line of Sec.~\ref{sec:anchor}, not the
boundary of \eqref{eq:syst}: to
its left the register's uncorrected systematic is more than a tenth of the signal,
$\rhob S<10\sigma_{\mathrm{s}}=1.7\times10^{-1}$. The existence boundary of
\eqref{eq:syst} sits a further factor ten to the left, at
$\rhob S=\sqrt{\lambda_{\mathrm{req}}}\sigma_{\mathrm{s}}=8.5\times10^{-2}$:
below it the corrected floor does not exist and no number of shots resolves the
bias, while between the two lines the verdict exists in principle but is
dominated by the systematic. The configurations here sit two decades to the left
of both, which is the sense in which
the shot budget is not the binding constraint. The two remedies are then
distinguishable at a glance: moving a point upward is a purchase, moving it
rightward is a calibration.

\section{External consistency check: a measured three-qubit register}
\label{sec:anchor}

Every number above is conditional on a model in which the only error is a coherent
generator and every gate is ideal. Three quantities inherit that assumption and can
be checked before any data are taken: the floor assumes unit assignment fidelity,
the residual assumes the coherent term dominates, and the margins between designs
assume a calibration stable at their size. A recent three-qubit experiment on the
same processor family supplies all three from measurement \cite{fezdata}. We use
only its device numbers---two calibration snapshots and the raw shot
records---and make no claim about its subject, which is unrelated to ours. The
register is qubits 0--2 of \texttt{ibm\_fez} (a Heron r2 processor---a
tunable-coupler device, hence exchange-family by the classification of
Sec.~\ref{sec:setting}; the block decomposition of Sec.~\ref{sec:coupling} was
computed for the dispersive platform, and on the exchange one the larger share
of the discrepancy, $98.1\%$, sits in the reference block, so this register's
verdict is expected to be the more design-blind kind), $8192$ shots
per setting, three circuits $\times$ six angles $\times$ three bases.

What this section is, and is not, should be said plainly. It is a consistency check
of the assumptions, not a demonstration of the protocol: no run of the sixteen
settings of Sec.~\ref{sec:protocol} exists, and nothing below is evidence that the
protocol works on hardware. What the check does have in its favour is independence.
The data come from a group that did not design this analysis, for a subject
unrelated to it, and neither the calibration nor the shot records could have been
tuned to make the model hold; a quantity that survives such a check has been
checked against something other than our own assumptions, which is what a
self-measurement on the same day cannot provide.

\emph{Readout.} The three assignment fidelities are $0.9844$, $0.9833$ and
$0.9954$; their product, $0.9634$, is the register's. Inverting the confusion
matrix inflates the variance of a component by $\prod_i f_i^{-2}=1.077$, so the
readout penalty on the floor is $7.7\%$: the estimate of Sec.~\ref{sec:scope} said
$6\%$, and the second snapshot gives a product of $0.9634$ again, so the number is
stable even though the individual fidelities are not. What the fidelity hides is
the asymmetry. The difference $p(1|0)-p(0|1)$ runs from $-0.39\%$ to $-1.95\%$
across the three sites and moved by $0.61\%$ on one site in three days. A
difference of settings, which is how the fingerprint is built, cancels that
asymmetry to first order; it does not cancel the gain. Uncured, the readout
multiplies every local angle by $\prod_i(1-a_i-b_i)=0.928$, a $7.2\%$ common
deficit. A common factor leaves the ranking of designs intact, and it is still not
a factor one may leave out of a quoted angle.

\emph{The floor is a systematic requirement before it is a shot count.} At the
floor, $1.07\times10^{7}$ shots, the per-component noise is $3.1\times10^{-4}$,
while the uncorrected readout bias on the worst site is $1.67\times10^{-2}$:
fifty-five times the statistical error. Shot noise stops being the binding term at
$3.6\times10^{3}$ shots, three thousand times below the floor, so every shot beyond
that buys nothing unless the readout has been characterized. Holding the model bias at a
tenth of the signal, $\rhob S=1.56\times10^{-3}$, requires it known to
$1.6\times10^{-4}$---a hundred times better than the chip delivers uncorrected.
That is the honest reading of item (iii) of Sec.~\ref{sec:scope}: the floor is what
the verdict costs given a readout calibration an order of magnitude better than the
one on the device. In the notation of \eqref{eq:syst},
$\sigma_{\mathrm{s}}=1.67\times10^{-2}$ gives
$N_{\mathrm{c}}=\sigma_{\mathrm{s}}^{-2}=3.6\times10^{3}$, three orders below
$\Nstar_{0}=1.07\times10^{7}$, so the denominator of the corrected floor is
negative: at this readout control there is no budget at which the test is
shot-limited, and the number this section reports is the upper bound, not the floor.
The statement can be made exact. As $N\to\infty$ the non-centrality of the test
saturates at $(\rhob S)^{2}/\sigma_{\mathrm{s}}^{2}=8.8\times10^{-3}$, which gives a
power of $0.050$---the size of the test, not a fraction more. Unlimited shots buy
nothing here at all.

\emph{Incoherent budget.} The model carries no gate error, so the coherent term is
the residual only if the stochastic weight of the circuit is smaller than
$\rhob$. The measured two-qubit errors are $1.64\%$ and $3.62\%$. A depolarising
layer is unital: it scales the traceless part of the transfer matrix uniformly, so
eight entangling gates map the generator to $(1-0.0263)^{8}=0.808$ of itself and every local
angle is $19\%$ small. Both families are linear subspaces of that traceless part, so the
fit, the residual of either family and their ordering all scale with it: the
verdict is unchanged and the floor is $1/0.808^{2}=1.53$ times larger, which is the
entire cost of the gate error. An error that shifts the generator rather than
scaling it does not have this property---that is the non-unital case of
\eqref{eq:syst}, and readout is the instance of it measured below. Keeping the
coherent term dominant requires $e_{2}\lesssim\rhob/8=5.2\times10^{-4}$, thirty-one
to sixty-nine times below the measured device. That is the regime boundary, stated
as a requirement rather than assumed.

\emph{Margins.} The one design margin the paper quotes is $0.7\%$, four gates
against three. The readout asymmetry is $1.95\%$ and the calibration moved by
$0.61\%$ in three days; both are larger. A miscalibration common to all sixty-four
designs cancels in the comparison, which is why the ranking of
Sec.~\ref{sec:design} survives it, but a sub-percent margin cannot be resolved
across calibration epochs: any claim at that level needs the readout characterized
in the same run that takes the data.

\emph{Two null tests.} Two of the recorded settings have an answer known before the
data are taken, and both return it. At zero rotation the teleported state is
$|0\rangle$, so every recorded $|1\rangle$ is an error: the run returns
$p_1=5.26\%$, the assignment errors of the three sites account for $0.27$ to
$1.25$ points of that, and once each site's gain is divided out $4.2$--$5.0\%$ is
left over, which is circuit and not
readout. That residue is $11\times$ the $\rhob$ this paper resolves---the
incoherent budget above, measured instead of inferred. With feedforward disabled
the correction is absent and the protocol must return the unpolarized value: it
returns $49.26\pm0.55\%$, $1.3\sigma$ from $50\%$, against $5.26\%$ with the
correction on, $73$ standard errors apart. Neither setting exercises the
fingerprint. What they establish is the narrower premise this section needs: the
counts move with a knob whose effect is known in advance, by the amount the model
predicts.

What device data can settle in advance, they
settle here, and all three answers are conservative: the readout penalty is $7.7\%$
rather than $6\%$, the floor is a systematic requirement before it is a shot
budget, and the family question is askable on a register whose two-qubit error is a
few times $10^{-4}$, not on the $10^{-2}$ hardware that exists.

\begin{table*}[t]
\caption{\label{tab:anchor}Three assumptions of the analysis against a measured
three-qubit register (qubits 0--2 of \texttt{ibm\_fez}, $8192$ shots per setting,
calibration 2026-06-14 and 2026-06-17, Ref.~\cite{fezdata}). The last column is the
consequence for the numbers of this paper. ``Bias'' is the symmetric readout error
of the worst site, $(\bar a+\bar b)/2$; the noise is the per-component
$1/\sqrt{N}$ at the floor $1.07\times10^{7}$.}
\small
\setlength{\tabcolsep}{7pt}
\begin{tabular}{llll}
\toprule
quantity & model & measured & consequence\\
\midrule
register assignment fidelity $\prod_i f_i$ & $1$ & $0.9634$ & floor $\times1.077$ (estimated $1.06$)\\
readout gain $\prod_i(1-a_i-b_i)$ & $1$ & $0.9277$ & every angle $7.2\%$ low, uncorrected\\
readout asymmetry $\max_i|a_i-b_i|$ & $0$ & $1.95\%$ & above the $0.7\%$ design margin\\
drift of that asymmetry over three days & $0$ & $0.61\%$ & above the $0.7\%$ margin\\
bias over noise at the floor & $0$ & $55$ & floor presumes a $1.6\times10^{-4}$ correction\\
shots where bias overtakes noise & --- & $3.6\times10^{3}$ & $3\times10^{3}$ below the floor\\
two-qubit error $e_2$ (CZ) & $0$ & $1.64\%$, $3.62\%$ & damping $0.808$; requires $5.2\times10^{-4}$\\
\bottomrule
\end{tabular}
\end{table*}

\section{Implications for reporting}
\label{sec:report}

A tomography report that quotes a residual, a gate count and a decoded logical
error has quoted three numbers whose domains are unstated.

With the residual go the shot floor and the design it was priced at, and the
floor is a systematic requirement before it is a shot count
(Sec.~\ref{sec:anchor}).
$1.1\times10^{7}$ and $6.7\times10^{6}$ shots per component buy $80\%$ power at
$\alpha=0.05$; at $10^{5}$ the residual reported is $5\%$ and the power equals the
level, so the family decision is unaffordable at the budget the protocol assumes,
and the honest entry in the report is ``undecided'' rather than a family. And the
floor belongs to the pair: the same device at the same working point spans
$1.44\times10^{6}$ to $9.74\times10^{7}$ across designs, so a floor quoted without
its design is the floor of an unstated experiment.

With the gate count go the placement and its spread. At three gates the rms
varies by a factor $2.7$--$3.9$ with placement alone; the optimal count is four
on the dispersive platform and three on the exchange one; the reference design is
the global optimum in one family and $0.7\%$ off it in the other. And the
placement is itself only weakly transferable: over forty working points the two
platforms agree on the argmin pair at $22$--$78\%$ of them, though choosing the
other platform's pair costs a median of at most $1.05$ and a ninetieth
percentile of $1.4$ above two gates. A count without
a placement is half a design, and the placement is not a constant either.

With a comparison between families go the normalisation, the design and the
strength convention. The claim holds under all seven normalisations at the
three-gate design, and under four of seven at the six-gate decoder at the
infidelity-matched strength---six at equal coupling angles---where the per-bond
coupling error is $1.46$ and $1.20$ times larger; without all three it is a
comparison of conventions rather than of archetypes. The per-bond entry is the
one that no convention rescues.

All three change a recommendation, not merely a caveat: the first says that the
family decision cannot be made at the assumed budget and that its price is a
design variable; the second that the design decision is not summarised by its
cost; the third that the family ordering is not summarised by a single number.

The limitations are those of the model, not of the method. The device is a
three-qubit chain and the families are the two archetypes above; a third family
would need its own $\nu$ and its own bias, and the criterion of
Sec.~\ref{sec:detect} applies to it unchanged. The residual test is asymptotic in
$\nu$ and linear-model in its degrees of freedom, both stated in
Sec.~\ref{sec:scope}. The Pareto statements are at $N=10^{5}$ and in the nine
local angles, and the objective is a further conditioning we can only report rather
than remove. Taking the coupling rms instead, design $A$ sits on the front in
neither family and $1.2\times$ (exchange) or $1.9\times$ (dispersive) above the
optimum, which is a different placement of the same three gates on the exchange
platform and a four-gate design on the dispersive one. That is the same lesson a
third time,
and it is the reason Sec.~\ref{sec:protocol} asks for the spread over placements
rather than for a winner. The design sweep of Sec.~\ref{sec:coupling} treats a
design as the recovery key of Sec.~\ref{sec:design}: an ideal correction applied
conditionally on the syndrome, with no gate error of its own. A correction gate
that is itself faulty would add a term to both fingerprints, and the direction of
that term is not predicted here. The working-point scans are samples---sixteen
points for the shot floor, forty for the design---not bounds, the first because
each point costs two multi-start fits. The block share is reported as two
readings at two register sizes, not as a scatter of share against design
sensitivity: that scatter needs the share resolved for every design of a third
register, and two sizes of one mechanism bound the mechanism here in a way a
larger sample at one size would not. The five-qubit spreads---the factor $119$
in rms and the factor $21$ in the verdict price---are properties of the six
designs sampled at that size, and a different sample would move them; they
illustrate the rule rather than measure it a second time.
And we compare two designs and seven
normalisations: a reader who prefers an eighth can compute it from the deposited
covariances.

\section{Outlook: more families, larger registers, other codes}
\label{sec:outlook}

Three generalisations are natural. The results above say how much of each can
be argued and how much has to be computed.

\emph{A cluster of families.} With $K$ candidates the test is run $K$ times, and a
family-wise level $\alpha$ costs a per-test level $\alpha/K$. The floor rises by
the ratio $\lambda_{\mathrm{req}}(\alpha/K)/\lambda_{\mathrm{req}}(\alpha)$, which
is logarithmic: $1.15$ for two families, $1.29$ for four, $1.54$ for sixteen,
$1.77$ for sixty-four. Widening the hypothesis space is therefore cheap next to
the other two handles---the working point moves the floor over four to five
decades, and the design by a factor $68$. A study that enlarges the model set
without choosing the design is spending its effort in the cheaper place. The
correction is the Bonferroni one and so conservative; what carries over is the
scaling, which is the same $\sqrt{\nu}$ behaviour that \eqref{eq:Nstar} inherits.

\emph{Register size.} We ran the same calculation at five qubits, where the
design space stops being trivial: $63$ effective designs per code instead of $8$
(the order cap is no longer redundant) and $3969$ pairs instead of $64$, of which
$3309$ are full rank on the dispersive platform and $3780$ on the exchange
one---so feasibility returns as a constraint, consistent with the note's exact
count $B^{*}=1$. The design's grip grows, and the mechanism says why. At five
qubits $71.8\%$ of the discrepancy sits in the phase code block, which a design
reshapes, and only $27.8\%$ in the reference block, which no design touches; at
three qubits the split is $52.0\%$ and $48.0\%$. The larger code has moved the
misspecification into the block the design can reach. Correspondingly the
placement spread at a fixed count rises from $2.7$
to $119$ at three gates, and the spread of the verdict price over the same sample
from $9.1$ to $21$ ($2.2\times10^{10}$ against $1.03\times10^{9}$ shots per
component, at similar $S$). The sample is the count-optimal placement at one, two, three,
four and six gates together with the full design---six designs here, seven at
three qubits, where the zero-gate design is also included; the rank and block
decompositions at five qubits are enumerated, the count scan is sampled. What does
not survive is the verdict. The
floor at the full design moves from $1.07\times10^{7}$ to $1.47\times10^{9}$
shots per component and reaches $2.2\times10^{10}$ over the sample, while the
local-angle rms runs from $1.3\times10^{4}$ to $3.2\times10^{5}$~mrad; at that
size no design makes the test affordable, and the design question stops being
the one that decides the experiment. Two caveats attach. The five-qubit fit is
sloppy in the technical sense of the word---a model manifold whose widths fall off
geometrically, so that the cost has long narrow canyons and a fit can stall short
of the minimum \cite{sloppy}---on the dispersive platform the fitted family's
Jacobian has condition number $1.8\times10^{5}$, and the residual falls from
$2.4\times10^{-4}$ to $5.8\times10^{-5}$ when the evaluation cap is raised eight
fold---so its floors
are lower bounds at a stated budget, and the design space there is sampled rather
than enumerated. The direction, however,
is the one the mechanism predicts: the reach grows where the code is, and at five
qubits the code is where the discrepancy is.

\emph{Other codes.} We have not computed a second code, and quote no number for
one; Sec.~\ref{sec:coupling} says what would decide the question: a code enters the
fingerprint through its block structure, so it re-prices the verdict only insofar as
the misspecification reaches the blocks the code adds. Our two cells are the extremes---$98.1\%$ of the exchange
discrepancy sits in the single-qubit block, which no code exposes better, $52.0\%$ of
the dispersive one in code blocks, where a design change moves the floor by $68$. The
prediction is falsifiable: a code that adds blocks without adding discrepancy cannot
lower the floor, which falls only when the code redistributes the discrepancy into
the blocks it creates. The five-qubit run is a first instance and it passes; a second
code would need the same number, the block decomposition of $\delta$.

\section{Data and code}
\label{sec:data}

Every number is reproduced by fourteen scripts and a shared library in the
deposit: \texttt{nm\_01}
(residuals, shot floor, operating characteristics, working-point scan),
\texttt{nm\_02} (the seven normalisations), \texttt{nm\_03} (the enumeration, the
Pareto fronts and the placement spreads), \texttt{nm\_04} (the design-resolved
fingerprints, the re-priced floor and the block decomposition of $\delta$),
\texttt{nm\_05} (the strength conventions and the sweep of the matching scale),
\texttt{nm\_06} (the family dependence of the argmin placement, at the reference
point and over forty sampled points), \texttt{nm\_07} (the gate-count note's
three-qubit table, recomputed at its own point, and the re-pricing there),
\texttt{nm\_08} (the same design rule at five qubits), \texttt{nm\_09} (the device
check of Sec.~\ref{sec:anchor}: readout budget, shot-versus-systematic crossover
and incoherent allowance, from the calibration and shot records of
Ref.~\cite{fezdata}), \texttt{nm\_10} (the check itself: it recomputes the
derivations, prints every headline number the way the text prints it and looks
for that string in the source, and compares a fresh run against the archived
one), \texttt{nm\_11} (the plane of Fig.~\ref{fig:phase}: it reads the bias and
the sensitivity of every configuration from the archived outputs rather than
refitting them, so the figure is a view of the tables and not a new
computation), \texttt{nm\_12} (equation~\eqref{eq:syst} in its five forms, and
the power the test saturates at), \texttt{nm\_13} (the bootstrap of
Sec.~\ref{sec:detect}(ii) at the reference point: synthetic counts, refits and
the size and power of $T$), \texttt{nm\_14} (the same bootstrap at the median
working point of each direction), and \texttt{nm\_lib} (the
shared definitions, including the design-resolved fingerprint in the generic
$n$-qubit form used by \texttt{nm\_08}). Their archived outputs are included, so
the tables can be checked
without running anything, and \texttt{nm\_10} fails loudly when a quoted number
and its stored value have drifted apart---the bootstrap sizes, powers and
corrected floors among them; \texttt{nm\_09} needs the deposited dataset of
Ref.~\cite{fezdata} in \texttt{device\_data/} and runs in seconds,
\texttt{nm\_13} and \texttt{nm\_14} take about half an hour each ($150$ refits
per experiment), \texttt{nm\_07} about five minutes,
\texttt{nm\_04} and \texttt{nm\_08} about four, \texttt{nm\_01}
about one, and the rest seconds to half a minute. Times are for Python 3.13
with NumPy 2.5 and SciPy 1.18 on eight cores; the versions are pinned in the
deposit's \texttt{requirements.txt}. The fingerprint, the design algebra and the
exact rank machinery are those of Refs.~\cite{note,supp}, deposited at Zenodo
(\href{https://doi.org/\zenodo}{doi:\zenodo}). This paper's own code and its
archived outputs are deposited separately at
\href{https://doi.org/\zenodocode}{doi:\zenodocode}; the two deposits are kept
apart because the earlier one is the companion note's supplement, not this
paper's archive. A third applies the same rule to fifteen effects that five
public datasets report, with none of this paper's data and no verdict on any
family: \href{https://doi.org/\zenodonote}{doi:\zenodonote}.

\begin{acknowledgments}
AI tools (DeepSeek-V4.1-Flash) were used to edit the language of the text and,
within specifications set by the author, to draft and debug parts of the analysis
code \texttt{nm\_01}--\texttt{nm\_14}. All code, output and wording were reviewed
and tested by the author; every derivation is rederived in closed form in the text,
and \texttt{nm\_10} compares a fresh run against the archived outputs ($6956$
floating-point values, zero relative difference) and looks for every quoted number
in the source as the text prints it.
\end{acknowledgments}

\renewcommand{\bibfont}{\footnotesize}


\begin{thebibliography}{9}
\bibitem{merkel} S. T. Merkel, J. M. Gambetta, J. A. Smolin, and B. R. Johnson,
``Self-consistent quantum process tomography,'' Phys. Rev. A \textbf{87}, 062119
(2013).
\bibitem{gst} R. Blume-Kohout, J. K. Gamble, E. Nielsen, K. Rudinger, J. Mizrahi,
K. Fortier, and P. Maunz, ``Demonstration of qubit operations below a rigorous
fault-tolerance threshold with gate set tomography,'' Nat. Commun. \textbf{8},
14485 (2017).
\bibitem{greenbaum} D. Greenbaum, ``Introduction to gate set tomography,''
arXiv:1509.02921.
\bibitem{dasilva} M. P. da Silva, O. Landon-Cardinal, and D. Poulin, ``Practical
characterization of quantum devices without tomography,'' Phys. Rev. Lett.
\textbf{107}, 210404 (2011).
\bibitem{flammia} S. T. Flammia and Y.-K. Liu, ``Direct fidelity estimation from
few Pauli measurements,'' Phys. Rev. Lett. \textbf{106}, 230501 (2011).
\bibitem{harper} R. Harper, S. T. Flammia, and J. J. Wallman, ``Efficient learning
of quantum noise,'' Nat. Phys. \textbf{16}, 1184 (2020).
\bibitem{eisert} J. Eisert, D. Hangleiter, N. Walk, I. Roth, D. Markham, R. Parekh,
U. Chabaud, and E. Kashefi, ``Quantum certification and benchmarking,'' Nat. Rev.
Phys. \textbf{2}, 382 (2020).
\bibitem{taxonomy} R. Blume-Kohout, M. P. da Silva, E. Nielsen, T. Proctor,
K. Rudinger, M. Sarovar, and K. Young, ``A taxonomy of small Markovian errors,''
PRX Quantum \textbf{3}, 020335 (2022).
\bibitem{burnham} K. P. Burnham and D. R. Anderson, \emph{Model Selection and
Multimodel Inference} (Springer, New York, 2002), 2nd ed.
\bibitem{note} L. Huang, ``Recovery gates as a resource in repetition-code process
tomography: exhaustive gate-optimal designs for three, five and seven qubits,''
arXiv:2611.08310 (2026); companion paper to the present one, the increment of
this paper over it being stated at the end of Sec.~\ref{sec:intro}.
\bibitem{main} L. Huang, ``Probing coherent noise with stabilizer codes:
identifiability, ambiguity, and decoder priors,'' arXiv:2610.19779 (2026).
\bibitem{supp} L. Huang, ``Experimental supplement: three-qubit, three-gate
repetition-code noise tomography,'' Zenodo, version 1.5 (2026),
\href{https://doi.org/\zenodo}{doi:\zenodo}.
\bibitem{nielsen} M. A. Nielsen, ``A simple formula for the average gate fidelity
of a quantum dynamical operation,'' Phys. Lett. A \textbf{303}, 249 (2002).
\bibitem{haah} J. Haah, A. W. Harrow, Z. Ji, X. Wu, and N. Yu, ``Sample-optimal
tomography of quantum states,'' IEEE Trans. Inf. Theory \textbf{63}, 5628--5641
(2017).
\bibitem{wallman} J. J. Wallman and S. T. Flammia, ``Randomized benchmarking with
confidence,'' New J. Phys. \textbf{16}, 103032 (2014).
\bibitem{kay} S. M. Kay, \emph{Fundamentals of Statistical Signal Processing,
Vol.~II: Detection Theory} (Prentice Hall, 1998).
\bibitem{sugiyama} T. Sugiyama, P. S. Turner, and M. Murao, ``Precision-guaranteed
quantum tomography,'' Phys. Rev. Lett. \textbf{111}, 160406 (2013).
\bibitem{pukelsheim} F. Pukelsheim, \emph{Optimal Design of Experiments} (SIAM,
Philadelphia, 2006), Classics in Applied Mathematics \textbf{50}.
\bibitem{kimmel} S. Kimmel, G. H. Low, and T. J. Yoder, ``Robust calibration of a
universal single-qubit gate set via robust phase estimation,'' Phys. Rev. A
\textbf{92}, 062315 (2015).
\bibitem{fezdata} V. Rai and S. Farhat, ``Entanglement enables coherence transfer and
feedforward corrects the state in quantum teleportation on superconducting dynamic
circuits,'' data and code, Zenodo (2026),
\href{https://doi.org/10.5281/zenodo.20767110}{doi:10.5281/zenodo.20767110}.
\bibitem{sloppy} M. K. Transtrum, B. B. Machta, and J. P. Sethna, ``Geometry of
nonlinear least squares with applications to sloppy models and optimization,''
Phys. Rev. E \textbf{83}, 036701 (2011).
\bibitem{cox} D. R. Cox, ``Tests of separate families of hypotheses,'' in
\emph{Proceedings of the Fourth Berkeley Symposium on Mathematical Statistics and
Probability}, Vol.~1 (University of California Press, Berkeley, 1961), p.~105.
\bibitem{vuong} Q. H. Vuong, ``Likelihood ratio tests for model selection and
non-nested hypotheses,'' Econometrica \textbf{57}, 307 (1989).
\end{thebibliography}
\end{document}